\documentclass[sn-mathphys,Numbered]{sn-jnl}

\usepackage{graphicx}
\usepackage{float}
\usepackage{multirow}%
\usepackage{amsmath,amssymb,amsfonts}
\usepackage{amsthm}
\usepackage{mathrsfs}
\usepackage[title]{appendix}
\usepackage{xcolor}
\usepackage{textcomp}
\usepackage{manyfoot}%
\usepackage{booktabs}%
\usepackage{algorithm}
\usepackage{mathtools}
\usepackage{algorithmicx}%
\usepackage{algpseudocode}%
\usepackage{listings}%
\usepackage{lmodern}
\usepackage[capitalize]{cleveref}
\usepackage{anyfontsize}
\usepackage{hyperref}
\usepackage{cleveref}
\usepackage{mathtools}

\begin{document}

\title{Dyonic Kerr-Sen Black Hole's Resonant Scattering: Absorption and Superradiance}

\author[1]{Supakorn Katewongveerachart}
\email{supakorn.katewongveerachart@gmail.com}
\author[1]{David Senjaya} \email{davidsenjaya@protonmail.com} 
\affil[1]{Department of Physics, Faculty of Science, Mahidol University\\ 272 Rama VI Road, Ratchathewi, Bangkok, Thailand}


\maketitle
\begin{abstract}

We analytically investigate scalar superradiant scattering in the rotating dyonic Kerr-Sen black hole of Einstein-Maxwell-Dilaton-Axion theory. Starting from the separable Klein-Gordon equation for a massive neutral scalar field, we work in the low-frequency and slow-rotation regime and employ the analytical asymptotic matching (AAM) method to compute the reflection coefficient and the associated superradiant amplification factor. Since an exact global scattering solution is not available in this four-charge geometry, the AAM framework enables a controlled analytic treatment of the near-and far-region dynamics. We provide detailed and systematic derivations of the matching procedure leading to the closed-form amplification formula. The superradiant condition is obtained explicitly and we demonstrate that energy extraction occurs exclusively for co-rotating modes satisfying $\Omega < m \Omega_H$. We show that the presence of electric and magnetic charges suppresses the amplification relative to the Kerr limit, whereas lighter co-rotating scalar fields broaden the superradiant window and enhance the efficiency of rotational energy extraction. 

\end{abstract}

\section{Introduction}
General relativity provides a geometric description of gravity in which spacetime curvature governs the motion of matter and radiation. Over the past century, its predictions have been confirmed with remarkable precision, including gravitational lensing, perihelion precession, gravitational redshift, gravitational waves and the existence of black holes \cite{Hobson,Will,Abbott,Abbott1,EHTC}. The recent era of precision black hole observations has further established these objects as direct probes of strong-field gravity.

Despite its empirical success, general relativity is widely expected to be incomplete. Classical solutions generically contain spacetime singularities where curvature invariants diverge and predictability breaks down. On galactic and cosmological scales, the theory requires dark matter and dark energy in the energy-momentum tensor $T_{\mu\nu}$ to account for rotation curves and cosmic acceleration. These challenges have motivated the development of extended gravitational theories, including $f(R)$ and $f(T)$ gravity, Gauss-Bonnet gravity, scalar-tensor-vector gravity (STVG), Lovelock gravity and Kaluza-Klein models \cite{Moffat_2006,MOFFAT_2007,Clifton_2012,Papantonopoulos:2015cva,Petrov_2020}.

Among such extensions, string-inspired models naturally introduce additional scalar and pseudo-scalar degrees of freedom. In this work we consider the Einstein-Maxwell-dilaton-axion (EMDA) theory, a scalar-vector-tensor framework emerging from low-energy supergravity. In EMDA gravity, the electromagnetic field couples nontrivially to a dilaton field $\xi$ and an axion field $\phi$, leading to a richer black hole structure than in Einstein-Maxwell theory. The four-dimensional effective action is given by \cite{Sen_1992,Banerjee_2020}
\begin{multline}
S_{EMDA}=\frac{1}{16\pi}\int\left[
R
-2\partial_\mu\xi\partial^\mu\xi
-\frac{1}{3}H_{\rho\sigma\delta}H^{\rho\sigma\delta}
+e^{-2\xi}F_{\alpha\beta}F^{\alpha\beta}
\right]\sqrt{-g}\, d^4x,
\end{multline}
where $R$ denotes the Ricci scalar and $g$ the determinant of the metric. The Maxwell field tensor is defined by
\begin{equation}
F_{\mu\nu}=\partial_\mu A_\nu-\partial_\nu A_\mu,
\end{equation}
while the Kalb-Ramond field strength is related to the axion through
\begin{equation}
H_{\alpha\beta\delta}
=\frac{1}{2}e^{4\xi}\varepsilon_{\alpha\beta\delta\gamma}
\partial^\gamma \phi.
\end{equation}

A significant development is the exact rotating four-charge solution obtained by Wu \cite{Wu_2021}, commonly known as the dyonic Kerr-Sen black hole. This geometry generalizes the original Kerr-Sen solution \cite{Sen_1992} by incorporating independent electric, magnetic, dilaton and axion charges. In the limit where the scalar sector vanishes, the solution reduces to the Kerr-Newman black hole of Einstein-Maxwell theory \cite{Kerr:1963ud}. The enlarged parameter space and nontrivial scalar couplings make this spacetime an ideal background for studying wave dynamics and spectral properties.

Black hole spectroscopy provides a powerful probe of spacetime geometry. Scalar perturbations obey the Klein-Gordon equation in curved spacetime and their characteristic frequencies encode geometric and dynamical information. In most rotating geometries, the radial equation does not admit closed-form solutions and one typically relies on numerical or semi-analytical techniques such as WKB approximations, asymptotic iteration methods, or continued fraction expansions. Although effective, these approaches often obscure the analytic structure underlying the quantization conditions.

In certain highly symmetric backgrounds, the separated radial equation can be expressed in terms of confluent or general Heun functions. In such cases, imposing the polynomial condition on the Heun function leads directly to discrete quantization rules for quasibound states. This analytic strategy has been successfully implemented for a variety of black hole spacetimes formulated in Boyer-Lindquist coordinates, including static dilatonic solutions \cite{Yang}, Einstein-Gauss-Bonnet black holes \cite{100}, Kerr-Newman geometries \cite{vier21,Senjaya:2025pyv}, Ernst black holes \cite{senjaya5}, EMDA configurations \cite{Senjaya:2024uqg,Senjaya:2025bbp,Senjaya:2024gpb}, Einstein-Bumblebee black holes \cite{Senjaya:2024blm,Senjaya:2025cgk} and Kerr-Einstein-Yang-Mills-Higgs solutions \cite{Senjaya:2024gfh}.

In particular, an exact analytic solution of the full radial scattering problem in the rotating dyonic Kerr-Sen geometry is impossible to be obtained. For this reason, we employ the analytical asymptotic matching (AAM) method \cite{Furuhashi,Hod:2013zza,Benone:2014ssa,Huang:2016qnk}, which provides a controlled analytic framework for extracting the superradiant amplification factor. The method takes advantage of the natural separation of scales in the spacetime. In the near-horizon region, where the black hole geometry dominates the dynamics, the radial equation simplifies and can be solved in terms of the hypergeometric function ${}_2F_1$, implementing the purely ingoing boundary condition at the event horizon. In the asymptotic region, where the spacetime approaches flatness and scattering states are well defined, the equation reduces to a confluent hypergeometric form described by ${}_1F_1$ functions. 

By analytically extending both solutions into their common overlap region and matching them consistently, the relative amplitudes of the ingoing and outgoing modes are uniquely determined. This procedure yields an explicit analytic expression for the superradiant amplification factor. The superradiant amplification factor for the rotating dyonic Kerr-Sen black hole has not been derived previously. Our analysis therefore provides the first analytic determination of the amplification coefficient in the most general configuration and clarifies how the electric, magnetic  charges together with the angular momentum influence the superradiant scattering process.

\section{The Dyonic Kerr–Sen Black Hole}
In this section we review the essential geometric properties of the Dyonic Kerr–Sen spacetime, determine its singular structure and horizons, construct the inverse metric explicitly and formulate the Klein–Gordon equation governing scalar perturbations.

We work in Boyer–Lindquist coordinates $(t,r,\theta,\phi)$, in which the Dyonic Kerr–Sen line element reads \cite{Wu,Jana},
\begin{multline}
ds^2 = -\left[1-\frac{r_s(r-d)-r_D^2}{\rho^2}\right]c^2 dt^2 -2\frac{r_s(r-d)-r_D^2}{\rho^2}a\sin^2\theta\, c\, dt\, d\phi \\ +\frac{\rho^2}{\Delta}dr^2 +\rho^2 d\theta^2 +\left[r(r-2d)-k^2+a^2 +\frac{r_s(r-d)-r_D^2}{\rho^2}a^2\sin^2\theta \right]\sin^2\theta\, d\phi^2,
\label{metric}
\end{multline}
where,
\begin{gather}
\rho^2 = r(r-2d)-k^2+a^2\cos^2\theta, \\
\Delta = r(r-2d)-r_s(r-d)-k^2+a^2+r_D^2.
\end{gather}

The parameters are defined by,
\begin{gather}
r_D^2 = \frac{G(Q_E^2+Q_M^2)}{4\pi\epsilon_0 c^4}=Q^2+P^2, \\
k = \frac{2PQ}{r_s}, \qquad
d = \frac{P^2-Q^2}{r_s}, \\
a = \frac{J}{M},
\end{gather}
where $M$, $J$, $Q$ and $P$ denote the mass, angular momentum, electric charge and magnetic charge of the black hole, while $d$ and $k$ encode the dilaton and axion contributions.

The condition $\Delta=0$ determines the horizons,
\begin{gather}
r_\pm = \frac{r_s}{2}+d \pm \sqrt{\left(\frac{r_s}{2}\right)^2+d^2+k^2-(a^2+r_D^2)},
\label{rpm}
\end{gather}
and surfaces $r=r_\pm$ correspond to the inner and outer horizons.

It is worth emphasizing that the Dyonic Kerr-Sen solution contains several familiar black hole geometries as limiting cases.  Setting $d=0$ in the definition of $k$ removes the dilatonic contribution and reproduces the non-dilatonic, axionic dyonic black hole. Conversely, suppressing the axionic sector yields the non-axionic, dilatonic dyonic configuration. If the magnetic charge is set to zero, $P=0$, the solution reduces to the Kerr-Newman black hole, which represents the most general stationary electro-vacuum solution of the Einstein-Maxwell theory. Imposing $P=0$ and $a=0$ further eliminates rotation and magnetic charge, leading to the static and spherically symmetric Reissner-Nordstrom solution. Finally, setting all parameters to zero except the mass reduces the geometry to the Schwarzschild black hole.

To construct the inverse metric of the Dyonic Kerr-Sen spacetime, we first rewrite equation \eqref{metric} in matrix form,
\begin{gather}
g_{\mu\nu}=
\begin{pmatrix}
-\left[1-\dfrac{r_s(r-d)-r_D^2}{\rho^2}\right] & 0 & 0 & g_{c\phi} \\
0 & \dfrac{\rho^2}{\Delta} & 0 & 0 \\
0 & 0 & \rho^2 & 0 \\
g_{c\phi} & 0 & 0 & g_{\phi\phi}
\end{pmatrix},
\end{gather}
where,
\begin{equation}
g_{c\phi}=
-\frac{r_s(r-d)-r_D^2}{\rho^2}\,a\sin^2\theta,
\end{equation}
\begin{multline}
g_{\phi\phi}=
\Bigg[
r(r-2d)-k^2+a^2 \\
+ \frac{r_s(r-d)-r_D^2}{\rho^2}a^2\sin^2\theta
\Bigg]\sin^2\theta .
\end{multline}

For computational convenience, we reorder the components so that the metric assumes a block-diagonal structure,
\begin{align}
g_{\mu\nu}
&=
\begin{pmatrix}
\dfrac{\rho^2}{\Delta} & 0 & 0 & 0 \\
0 & \rho^2 & 0 & 0 \\
0 & 0 &
-\left[1-\dfrac{r_s(r-d)-r_D^2}{\rho^2}\right] & g_{c\phi} \\
0 & 0 & g_{c\phi} & g_{\phi\phi}
\end{pmatrix}
\nonumber \\
&=
\begin{pmatrix}
\tilde A & 0 \\
0 & \tilde B
\end{pmatrix}.
\end{align}

The determinant follows immediately from the block structure,
\begin{equation}
g=\det(g_{\mu\nu})
=\det(\tilde A)\det(\tilde B).
\end{equation}

A straightforward evaluation yields,
\begin{equation}
g=-\rho^4\sin^2\theta.
\end{equation}

Because the metric has block form, its inverse can be obtained blockwise,
\begin{equation}
g^{\mu\nu}=
\begin{pmatrix}
\tilde A^{-1} & 0 \\
0 & \tilde B^{-1}
\end{pmatrix}.
\end{equation}

The first block is diagonal and inverts trivially,
\begin{equation}
\tilde A^{-1}=
\begin{pmatrix}
\dfrac{\Delta}{\rho^2} & 0 \\
0 & \dfrac{1}{\rho^2}
\end{pmatrix}.
\end{equation}

For the second block we obtain,
\begin{equation}
\tilde B^{-1}
=\frac{1}{|\tilde B|}
\begin{pmatrix}
g_{\phi\phi} & -g_{c\phi} \\
-g_{c\phi} &
-\left[1-\dfrac{r_s(r-d)-r_D^2}{\rho^2}\right]
\end{pmatrix},
\end{equation}
with,
\begin{equation}
|\tilde B|
=\frac{-g}{\rho^4/\Delta}
=-\Delta\sin^2\theta.
\end{equation}

Combining both blocks, the inverse metric takes the explicit form,
\begin{gather}
g^{\mu\nu}=
\begin{pmatrix}
-\dfrac{1}{\Delta}f_{\phi\phi} & 0 & 0 &
\dfrac{g_{c\phi}}{\Delta\sin^2\theta} \\
0 & \dfrac{\Delta}{\rho^2} & 0 & 0 \\
0 & 0 & \dfrac{1}{\rho^2} & 0 \\
\dfrac{g_{c\phi}}{\Delta\sin^2\theta} & 0 & 0 &
\dfrac{1}{\Delta\sin^2\theta}
\left(1-\dfrac{r_s(r-d)-r_D^2}{\rho^2}\right)
\end{pmatrix},
\\
f_{\phi\phi}
=
r(r-2d)-k^2+a^2
+\frac{r_s(r-d)-r_D^2}{\rho^2}a^2\sin^2\theta .
\end{gather}

\section{The Boson Dynamics}
The covariant Klein-Gordon equation in a curved spacetime is,
\begin{equation}
-\hbar^2 \nabla_\mu \nabla^\mu \psi + m^2 c^2 \psi = 0,
\end{equation}
where $\nabla_\mu$ denotes the covariant derivative compatible with the metric.

Using,
\[
\nabla_\mu \nabla^\mu \psi = \partial_\mu \partial^\mu \psi + \Gamma^\mu_{\mu\nu}\partial^\nu \psi,
\]
together with the identity,
\[
\Gamma^\alpha_{\alpha\beta} = \frac{1}{\sqrt{-g}} \frac{\partial \sqrt{-g}}{\partial x^\beta},
\]
the d'Alembert operator can be written in the compact Laplace-Beltrami form,
\begin{equation}
\nabla_\mu \nabla^\mu \psi = \frac{1}{\sqrt{-g}} \partial_\mu \left( \sqrt{-g}\, g^{\mu\nu}\partial_\nu \psi \right).
\end{equation}

The Klein-Gordon equation therefore becomes,
\begin{equation}
\left[ -\hbar^2 \frac{1}{\sqrt{-g}} \partial_\mu \left( \sqrt{-g}\, g^{\mu\nu}\partial_\nu \right) + m^2 c^2 \right]\psi = 0.
\end{equation}

Since both the determinant $g$ and the inverse metric $g^{\mu\nu}$ have already been determined, we now evaluate the Laplace-Beltrami operator component by component.
\begin{gather}
\frac{1}{\sqrt{-g}} \partial_0 \left( \sqrt{-g} g^{00} \partial_0 \right) = -\frac{1}{\Delta \rho^2} \left\{ \left[r(r-2d)-k^2+a^2\right]^2 - \Delta a^2 \sin^2\theta \right\} \partial_{ct}^2, \\ \frac{1}{\sqrt{-g}} \partial_0 \left( \sqrt{-g} g^{03} \partial_3 \right) = -\frac{\left[r(r-2d)-k^2+a^2-\Delta\right]a} {\Delta \rho^2} \partial_{ct}\partial_\phi, \\ \frac{1}{\sqrt{-g}} \partial_3 \left( \sqrt{-g} g^{30} \partial_0 \right) = -\frac{\left[r(r-2d)-k^2+a^2-\Delta\right]a} {\Delta \rho^2} \partial_{ct}\partial_\phi, \\ \frac{1}{\sqrt{-g}} \partial_1 \left( \sqrt{-g} g^{11} \partial_1 \right) = \frac{1}{\rho^2} \partial_r \left( \Delta \partial_r \right), \\ \frac{1}{\sqrt{-g}} \partial_2 \left( \sqrt{-g} g^{22} \partial_2 \right) = \frac{1}{\rho^2 \sin\theta} \partial_\theta \left( \sin\theta\, \partial_\theta \right), \\ \frac{1}{\sqrt{-g}} \partial_3 \left( \sqrt{-g} g^{33} \partial_3 \right) = \frac{\Delta - a^2 \sin^2\theta} {\Delta \sin^2\theta\, \rho^2} \partial_\phi^2.
\end{gather}

Collecting all contributions, the Klein-Gordon equation in the Dyonic Kerr-Sen background assumes the explicit form,
\begin{multline}
\Bigg[ -\frac{1}{\Delta \rho^2} \left\{ \left[r(r-2d)-k^2+a^2\right]^2 - \Delta a^2 \sin^2\theta \right\} \partial_{ct}^2 \\ -2\frac{\left[r(r-2d)-k^2+a^2-\Delta\right]a} {\Delta \rho^2} \partial_{ct}\partial_\phi + \frac{1}{\rho^2} \partial_r \left( \Delta \partial_r \right) \\ + \frac{1}{\rho^2 \sin\theta} \partial_\theta \left( \sin\theta\, \partial_\theta \right) + \frac{\Delta - a^2 \sin^2\theta} {\Delta \sin^2\theta\, \rho^2} \partial_\phi^2 \Bigg]\psi - \frac{m^2 c^2}{\hbar^2}\psi =0.
\label{fullwave}
\end{multline}

The temporal and azimuthal symmetries allow separation of variables using,
\begin{gather}
\psi(t,r,\theta,\phi) = e^{-iEt/\hbar + i m_l \phi} R(r) T(\theta).
\end{gather}

\subsection{The Polar Equation}
Define the dimensionless parameters,
\begin{gather}
\Omega = \frac{E r_s}{\hbar c},
\qquad
\Omega_0 = \frac{E_0 r_s}{\hbar c},
\end{gather}
where $E_0 = mc^2$.

Substituting into equation \eqref{fullwave} and multiplying by $r^2/\psi$ gives,
\begin{multline}
-\frac{1}{\Delta \rho^2} \left[ \left(r(r-2d)-k^2+a^2\right)^2 -\Delta a^2 \sin^2\theta \right] \left(-\frac{E^2}{\hbar^2 c^2}\right) \nonumber \\ -2\frac{\left(r(r-2d)-k^2+a^2-\Delta\right)a} {\Delta \rho^2} \left(\frac{E m_l}{\hbar c}\right) +\frac{1}{R\rho^2}\partial_r(\Delta \partial_r R) \nonumber \\  +\frac{1}{T\rho^2 \sin\theta} \partial_\theta(\sin\theta \partial_\theta T) +\frac{\Delta - a^2 \sin^2\theta} {\Delta \sin^2\theta \rho^2} (-m_l^2) -\frac{m^2 c^2}{\hbar^2} =0.
\end{multline}

Multiplying by $\rho^2$ and using $\sin^2\theta = 1-\cos^2\theta$,
\begin{multline}
\frac{1}{T \sin\theta} \partial_\theta(\sin\theta \partial_\theta T) -\frac{m_l^2}{\sin^2\theta} -\left( \frac{\Omega_0^2 a^2}{r_s^2} -\frac{\Omega^2 a^2}{r_s^2} \right) \cos^2\theta \nonumber \\ +\frac{1}{R}\partial_r(\Delta \partial_r R) +\frac{\Omega^2}{r_s^2} \frac{\left(r(r-2d)-k^2+a^2\right)^2}{\Delta} -\frac{\Omega^2 a^2}{r_s^2} \nonumber \\ -2\frac{\left(r(r-2d)-k^2+a^2-\Delta\right)a}{\Delta} \left(\frac{\Omega m_l}{r_s}\right) +\frac{m_l^2 a^2}{\Delta} -\frac{\Omega_0^2}{r_s^2} \left(r(r-2d)-k^2\right) =0.
\end{multline}

The angular equation becomes,
\begin{align}
\frac{1}{T \sin\theta} \partial_\theta(\sin\theta \partial_\theta T) -\frac{m_l^2}{\sin^2\theta} -\left( \frac{\Omega_0^2 a^2}{r_s^2} -\frac{\Omega^2 a^2}{r_s^2} \right) \cos^2\theta +\lambda_l^{m_l} =0.
\end{align}

In case of $a=k=0$, we recover $\lambda_l^{m_l}=l(l+1)$ and
$T(\theta)=P_l^{m_l}(\cos\theta)$. For $a\neq0$, the solution is spheroidal function,
\begin{equation}
T(\theta) = S_l^{m_l}(c_a,\cos\theta) = \sum_{r=-\infty}^{\infty} d_r^{l m_l}(c_a) P_{l+r}^{m_l}(\cos\theta),
\end{equation}
with,
\begin{equation}
c_a = \frac{\Omega_0^2 a^2}{r_s^2} - \frac{\Omega^2 a^2}{r_s^2}.
\end{equation}

\subsection{The Radial Equation}
After separating the angular part, the radial equation takes the form,
\begin{multline}
\partial_r\left(\Delta \partial_r R\right) +\Bigg[ \frac{\Omega^2}{r_s^2} \frac{\left(r(r-2d)-k^2+a^2\right)^2}{\Delta} -\frac{\Omega^2 a^2}{r_s^2} \\ -2\frac{\left(r(r-2d)-k^2+a^2-\Delta\right)a}{\Delta} \left(\frac{\Omega m_l}{r_s}\right) +\frac{m_l^2 a^2}{\Delta} \\ -\frac{\Omega_0^2}{r_s^2}\left(r(r-2d)-k^2\right) -\lambda_l^{m_l} \Bigg]R=0 .
\label{radialeq}
\end{multline}

Since the equation is singular at the roots of $\Delta=0$, corresponding to the inner and outer horizons $r_\pm$, we factorize,
\begin{equation}
\Delta=(r-r_-)(r-r_+),
\end{equation}
and define,
\begin{equation}
\delta_r \equiv r_+ - r_- .
\end{equation}

Using the identity,
\begin{equation}
\frac{\delta_r}{\Delta}=\frac{1}{r-r_+}-\frac{1}{r-r_-},
\end{equation}
the radial equation can be rearranged into the compact form,
\begin{multline}
\partial_r\left(\Delta \partial_r R\right) +\Bigg[ \frac{1}{\Delta} \left\{ \frac{\Omega}{r_s} \left(r(r-2d)-k^2+a^2\right) -m_l a \right\}^2 \\ - \left\{ \frac{\Omega_0^2}{r_s^2} \left(r(r-2d)-k^2+a^2\right) +K_l^{m_l} \right\} \Bigg]R=0 ,
\label{radialmod}
\end{multline}
where,
\begin{equation}
K_l^{m_l} = \frac{\Omega^2 a^2}{r_s^2} - \frac{\Omega_0^2 a^2}{r_s^2} - 2\frac{\Omega m_l}{r_s}a + \lambda_l^{m_l}.
\end{equation}

We now restrict attention to the exterior region,
\begin{equation}
r_+ \le r < \infty,
\end{equation}
and introduce the dimensionless variable,
\begin{equation}
\delta_r y = r-r_+, 
\qquad
r-r_- = \delta_r (y+1),
\end{equation}
which maps the domain to,
\begin{equation}
0 \le y < \infty .
\end{equation}

In terms of $y$, the radial equation becomes,
\begin{equation}
\partial_y^2 R + \left( \frac{1}{y} + \frac{1}{y+1} \right)\partial_y R + \left[F.T.^2 + S.T.\right]R =0 ,
\end{equation}
where the first term is defined as,
\begin{multline}
F.T. = \frac{1}{\delta_r} \left( \frac{1}{y} - \frac{1}{y+1} \right) \\ \times \left\{ \frac{\Omega}{r_s} \left[ (\delta_r y+r_+)^2 -2d(\delta_r y+r_+) -k^2+a^2 \right] - m_l a \right\}.
\end{multline}

After partial fraction decomposition, this simplifies to,
\begin{equation}
F.T. = \frac{\Omega}{r_s}\delta_r + \frac{K_1}{\delta_r y} + \frac{K_3}{y+1},
\end{equation}
with,
\begin{align}
K_1 &= \frac{\Omega}{r_s} \left[r_+(r_+-2d)-k^2+a^2\right] - m_l a, \\ K_3 &= \frac{\Omega}{r_s}(r_+ + r_- - 2d) - \frac{K_1}{\delta_r}.
\end{align}

Squaring yields,
\begin{multline}
F.T.^2 = \frac{\Omega^2}{r_s^2}\delta_r^2 + \frac{K_1^2}{\delta_r^2 y^2} + \frac{K_3^2}{(y+1)^2} \\ + \frac{2\Omega K_1}{r_s y} + \frac{2\Omega \delta_r K_3}{r_s (y+1)} + \frac{2K_1K_3}{\delta_r} \left( \frac{1}{y} - \frac{1}{y+1} \right).
\end{multline}

Finally, the fully decomposed radial equation assumes the form,
\begin{multline}
\partial_y^2 R + \left( \frac{1}{y} + \frac{1}{y+1} \right)\partial_y R \\ + \Bigg[ \left( \frac{\Omega^2}{r_s^2} - \frac{\Omega_0^2}{r_s^2} \right)\delta_r^2 + \frac{A_1}{y} + \frac{A_2}{y+1} + \frac{K_1^2}{\delta_r^2 y^2} + \frac{K_3^2}{(y+1)^2} \Bigg]R =0 ,
\label{radialeqy}
\end{multline}
where $A_1$ and $A_2$ collect the linear coefficients in $1/y$ and $1/(y+1)$.

\section{Resonant Boson Scattering}
Let us now reconsider \eqref{radialeq} and introduce the tortoise coordinate defined by,
\begin{equation}
    dr_* = \frac{r(r-2d) - k^2 + a^2}{\Delta}\, dr
    = \frac{\Sigma}{\Delta}\, dr ,
\end{equation}
together with the redefinition of the radial function,
\begin{equation}
    R(r) = \frac{R^{\ast}(r^{\ast})}{\sqrt{\Sigma}} .
\end{equation}

With these transformations, the radial equation takes a Schrödinger–like form,
\begin{multline}
\frac{\partial^{2} R^{\ast}}{\partial r^{\ast 2}} + \Bigg[ \frac{1}{\Sigma^{2}} \left( \frac{\Sigma \Omega}{r_{s}} - \frac{m_{l} a}{r_{s}} \right)^{2} - \frac{\Delta}{\Sigma^{2}} \left( \lambda_l^{m_l} + \frac{(r^{2}-2dr-k^{2})\Omega_{0}^{2}}{r_{s}^{2}} + \frac{\Omega^{2}}{r_{s}^{2}} a^{2} - \frac{2\Omega}{r_{s}} m_{l} a \right) \\ + \frac{\Delta^{2}}{\Sigma^{4}} \left( 2r^{2} - 4dr + k^{2} + 3d^{2} - a^{2} \right) + \frac{\Delta\,\partial_{r}\Delta}{\Sigma^{3}} (d - r) \Bigg] R^{\ast} = 0 .
\end{multline}

In the asymptotic regions, the effective potential approaches constant values. At spatial infinity,
\begin{equation}
r \to \infty \;\Longleftrightarrow\; r^{\ast} \to \infty,
\end{equation}
the potential reduces to
\begin{equation}
V_{\mathrm{eff}}(r) \approx \frac{\Omega^{2}}{r_{s}^{2}} - \frac{\Omega_{0}^{2}}{r_{s}^{2}} \equiv k_{\infty}^{2},
\end{equation}
so that the asymptotic wave number is
\begin{equation}
k_{\infty} = \frac{\sqrt{\Omega^{2}-\Omega_{0}^{2}}}{r_{s}} .
\end{equation}

Near the outer horizon,
\begin{equation}
r \to r_{+} \;\Longleftrightarrow\; r^{\ast} \to -\infty,
\end{equation}
the potential again approaches a constant,
\begin{equation}
V_{\mathrm{eff}}(r_{+}) \approx \left( \frac{\Omega}{r_{s}} - \frac{m_{l} a}{r_{s}\,\Sigma(r_{+})} \right)^{2} \equiv k_{+}^{2},
\end{equation}
which defines the near–horizon wave number,
\begin{equation}
k_{+} = \frac{\Omega - \Omega_{H}}{r_{s}},
\end{equation}
where the critical frequency is given by,
\begin{equation}
\frac{\Omega_{H}}{r_{s}}= \frac{m_{l} a}{r_{s}\,\Sigma(r_{+})}.
\end{equation}

The radial current associated with the conserved flux is defined as,
\begin{equation}
J = \mathrm{Re} \left( - i \, R^{\ast \dagger} \, \partial_{r^{\ast}} R^{\ast} \right).
\end{equation}

At spatial infinity, the solution takes the plane–wave form,
\begin{equation}
R_{\infty}^{\ast} = R_{\mathrm{inc}} e^{-i k_{\infty} r^{\ast}} + R_{\mathrm{ref}} e^{i k_{\infty} r^{\ast}}, \label{Rastinf}
\end{equation}
leading to the asymptotic current,
\begin{equation}
J_{\infty}=k_{\infty}\left(|R_{\mathrm{ref}}|^{2}-|R_{\mathrm{inc}}|^{2}\right).
\end{equation}

Near the horizon, regularity requires a purely ingoing solution,
\begin{equation}
R_{+}^{\ast} = R_{\mathrm{trans}} e^{-i k_{+} r^{\ast}},
\end{equation}
with corresponding flux,
\begin{equation}
J_{+} = - k_{+} |R_{\mathrm{trans}}|^{2}.
\end{equation}

Imposing flux conservation, $J_{\infty} = J_{+}$, yields,
\begin{equation}
|R_{\mathrm{ref}}|^{2}=|R_{\mathrm{inc}}|^{2}-\frac{k_{+}}{k_{\infty}}|R_{\mathrm{trans}}|^{2}.
\end{equation}

Substituting the explicit expressions for the wave numbers,
\begin{equation}
k_{\infty} = \frac{\sqrt{\Omega^{2}-\Omega_{0}^{2}}}{r_{s}}, \qquad k_{+} = \frac{\Omega - \Omega_{H}}{r_{s}},
\end{equation}
we finally obtain,
\begin{equation}
|R_{\mathrm{ref}}|^{2} = |R_{\mathrm{inc}}|^{2} - \frac{\Omega - \Omega_{H}} {\sqrt{\Omega^{2}-\Omega_{0}^{2}}} |R_{\mathrm{trans}}|^{2}, 
\end{equation}
or in terms of reflection $|\mathcal{R}|^{2}$ and transmission probability $|\mathcal{T}|^{2}$ as follows,
\begin{equation}
|\mathcal{R}|^{2} =1 - \frac{\Omega - \Omega_{H}} {\sqrt{\Omega^{2}-\Omega_{0}^{2}}} |\mathcal{T}|^{2}, \label{Zori}
\end{equation}

The superradiant behaviour is immediately determined by the sign of $k_{+}$. Since $k_{\infty} > 0$ for propagating modes with $\Omega > \Omega_0$, the reflected amplitude satisfies,
\begin{equation}
|R_{\mathrm{ref}}|^{2} > |R_{\mathrm{inc}}|^{2} \qquad \text{if and only if} \qquad k_{+} < 0 .
\end{equation}

Using $k_{+} = (\Omega - \Omega_c)/r_s$, this condition becomes,
\begin{equation}
\Omega < \Omega_c .
\end{equation}

In this regime the horizon flux is negative, indicating that rotational energy is extracted from the black hole and the reflected wave is amplified. For $\Omega > \Omega_c$, the flux is inward and ordinary absorption occurs.

It is important to emphasize that the above argument only determines the \emph{threshold} of superradiance through the sign of $k_{+}$. The analysis based on the tortoise coordinate and near-horizon flux conservation identifies the cutoff frequency $\Omega_c$, but it does not provide the explicit dependence of the reflection and transmission coefficients on $\Omega$. In particular, it does not yield the amplification factor as a function of frequency. To find analytical expression of the amplification factor or greybody factor and the frequency-dependent superradiant amplification, one must determine the ratio $R_{\mathrm{ref}}/R_{\mathrm{inc}}$ explicitly.

\section{Analytical Asymptotic Matching}\label{sect:insta}
In this section, we derive analytically the wave amplification coefficient of scalar fields propagating in the dyonic Kerr-Sen black hole background using the method of asymptotic matching (AAM). The essence of the approach is to exploit the natural structure of the spacetime by studying the wave equation in two physically distinct regions, i.e. close to the event horizon and far away from the black hole. Near the horizon, the geometry dictates the dominant behavior of the scalar field and the radial equation simplifies to a form that can be solved in terms of ${}_2F_1$ hypergeometric functions. Far from the black hole, the spacetime asymptotically flattens and the radial equation again reduces to a tractable form, admitting solutions of the ${}_1F_1$ hypergeometric function class.

The key of the AAM method is to connect these two regimes. By using the connection formulas of the hypergeometric functions, the near-horizon solution is analytically extended toward large $r$ while the asymptotically far solution is analytically extended toward the horizon $r_+$. Both the near and far solutions are then matched in a common intermediate region where they overlap. Matching them in this domain fixes the relative amplitudes of the ingoing and outgoing modes, allowing the wave amplification coefficient to be determined explicitly.

Let us start with the radial Klein-Gordon equation in the dyonic Kerr-Sen black hole spacetime \eqref{radialeqy}.  Multiplying the radial equation by $y^2(y+1)^2$, we obtain,
\begin{align}
0 &= y^2(y+1)^2{\partial }^{2}_yR+P\left(y\right){\partial }_yR+Q\left(y\right)R, \label{modrad}
\end{align}
where,
\begin{equation}
    P(y)=y + 3y^2 + 2y^3,
\end{equation}
\begin{multline}
    Q(y)=\frac{K_1^2}{\delta_r^2} + y \left[ \frac{2K_1 r_s (K_1 + K_3 \delta_r) - K_2 r_s \delta_r^2 + 2K_1 \delta_r^2 \Omega} {r_s \delta_r^2} \right] \\+ y^2 \left[ \frac{r_s^2 (K_1 + K_3 \delta_r)^2 + r_s^2 \delta_r^2 (K_4 - 2K_2) + 2 r_s \delta_r^2 \Omega (2K_1 + K_3 \delta_r) + \delta_r^4 (\Omega^2 - \Omega_0^2)} {r_s^2 \delta_r^2} \right] \\- y^3 \left[ \frac{r_s^2 (K_2 - K_4) - 2 r_s \Omega (K_1 + K_3 \delta_r) - 2 \delta_r^2 (\Omega^2 - \Omega_0^2)} {r_s^2} \right] + y^4 \left[ \frac{\delta_r^2 (\Omega^2 - \Omega_0^2)} {r_s^2} \right].
\end{multline}

\subsection{Far Field Limit}
Let us consider the radial equation \eqref{modrad} and take the limit $z\to\infty$ with $|a|, |P|, |Q|\ll 1$ and $\lambda_l^{m_l}\approx l(l+1)$, we find,
\begin{gather}
{\partial }^{2}_yR+\frac{2}{y}\partial_yR+\left[A_0+\frac{A_1}{y}+\frac{A_2}{y^2}\right] R=0, \label{farfieldequation1}\\
A_0=(\Omega^2 - \Omega_0^2) \left[1 - \frac{4 (P^2 + Q^2)}{r_s^2}\right],\\
A_1=\left(4\Omega^2 - 3\Omega_0^2\right) - \frac{2}{r_s^2} \left[ 6 (P^2 + Q^2)\Omega^2 - 5 (P^2 + Q^2)\Omega_0^2 \right],\\
A_2=-l(l+1) + 6\Omega^2 - 3\Omega_0^2 +\frac{  4 (P^2 + Q^2)\left(2\Omega_0^2-3\Omega^2 \right) }{r_s^2}. \label{A2far}
\end{gather}

By following Appendix \ref{AppendixA}, the far-field equation \eqref{farfieldequation1} can be written in the normal form,
\begin{gather}
\partial_y^2\mathcal{R} + \left[ A_0 + \frac{A_1}{y} + \frac{A_2}{y^2} \right] \mathcal{R} = 0,
\end{gather}
where we have defined the rescaled function,
\begin{gather}
\mathcal{R} = y R.
\end{gather}

Introducing the new radial variable,
\begin{gather}
z = 2 i \sqrt{A_0} \, y,
\end{gather}
the equation becomes,
\begin{gather}
\partial_z^2\mathcal{R} + \left[ -\frac{1}{4} + \frac{A_1}{2 i \sqrt{A_0} \, z} + \frac{A_2}{z^2} \right] \mathcal{R} = 0.
\label{normalformfarfieldequation1}
\end{gather}

Equation \eqref{normalformfarfieldequation1} is now in the standard confluent hypergeometric form. The general solution can therefore be expressed in terms of the confluent hypergeometric functions ${}_1F_1(a,b,z)$ (see Appendix \ref{AppendixC}),
\begin{multline}
R_{\text{far}} = N_{\infty 1} e^{-\frac{1}{2} i \alpha x} x^{\beta-\frac{1}{2}} F_{1,1} \left( \frac{1}{2}+\beta+i\frac{A_1}{\alpha}, 1+2\beta, i\alpha x \right)  \\ + N_{\infty 2} e^{-\frac{1}{2} i \alpha x} x^{-\beta-\frac{1}{2}} F_{1,1} \left( \frac{1}{2}-\beta+i\frac{A_1}{\alpha}, 1-2\beta, i\alpha x \right),
\label{farsol}
\end{multline}
where,
\begin{align}
\alpha &= 2\sqrt{A_0}, \\
\beta &= \frac{1}{2}\sqrt{1-4A_2}.
\end{align}

The integration constants are denoted by $N_{\infty 1}$ and $N_{\infty 2}$.  

Taking the limit $x \to 0$ and using $F_{1,1}(a,b,0)=1$, the far-field solution reduces to,
\begin{equation}
R_{\text{far}}(x\to 0) \approx N_{\infty 1} x^{\beta-\frac{1}{2}} + N_{\infty 2} x^{-\beta-\frac{1}{2}}.
\end{equation}

\subsection{Near Horizon Solution}
Let us start with the radial Klein-Gordon equation in the dyonic Kerr-Sen black hole spacetime \eqref{radialeqy}.  Multiplying the radial equation by $y(y+1)$ and considering the case in the vicinity of the event horizon, corresponding to the limit $z \to 0$, the radial equation reduces to,
\begin{gather}
y (y+1)\partial_y^2 R + (1+2y) \partial_y R + \left( B_0+\frac{B_1}{y} - \frac{B_2}{y+1} \right) R = 0,\label{10}
\\
B_0=-l(l+1) + 6\Omega^2 - 3\Omega_0^2+\frac{  4 (P^2 + Q^2)\left(2\Omega_0^2-3\Omega^2 \right) }{r_s^2}\\
B_1 =\frac{a^{2} m_l^{2}}{r_{s}^{2}}
- \frac{2 a m_l \left( 2\left(P^{2}+Q^{2}\right) + r_{s}^{2} \right)\Omega}{r_{s}^{3}}
+ \Omega^{2},\\
B_2 =3 \Omega^2 - 2 \Omega_0^2 +\frac{ 2 (P^2 + Q^2)( 3\Omega_0^2-4 \Omega^2  )}{ r_s^2 }.
\label{B2near}
\end{gather}

Interestingly, we recognize that $B_1$ is directly connected to the black hole's event horizon's angular momentum $\Omega_H$ as follows,
\begin{equation}
    \Omega_H=\frac{a r_s}{r_+(r_+-2d)+a^2-k^2},
\end{equation}
where,
\begin{equation}
 B_1 =\lim_{a,Q,P\to 0} \left(\Omega-m_l \Omega_H\right)^2.
\end{equation}

Clearly, static black holes, with $a=0$, has $\Omega_H=0$.

By comparing the near horizon equation \eqref{10} with \eqref{GaussModified}, we obtain the following ingoing wave solution,
\begin{equation}
       R_{near}=N_{H1}(y+1)^{-i\sqrt
       B_2}y^{-i\sqrt{B_1}} {}_2F_{1}\left(a_1,a_2,a_3,-y\right),
\end{equation}
where,
\begin{align}
    a_1&=\frac{1}{2}+\frac{1}{2}\sqrt{1-4B_0}-i\left(\sqrt{B_1}+\sqrt{B_2}\right),\\
    a_2&=\frac{1}{2}-\frac{1}{2}\sqrt{1-4B_0}-i\left(\sqrt{B_1}+\sqrt{B_2}\right),\\
    a_3&=1-2i\sqrt{B_1}.
\end{align}

In the overlap region, i.e., $R_{far}\left(y\to 0\right)=R_{near}\left(y\to \infty \right)$ the solutions are matched as follows,
\begin{gather}
R_{far}\left(y\to 0\right)\approx N_{\infty 1}y^{\beta-\frac{1}{2}}+N_{\infty 2}y^{-\beta-\frac{1}{2}},\\
 R_{near}\left(y\to \infty \right)\approx N_{H1}y^{-i\left(\sqrt{B_1}+\sqrt{B_2}\right)}\left[y^{-a_1}\frac{\Gamma \left(a_3\right)\Gamma \left(a_2-a_1\right)}{\Gamma \left(a_3-a_1\right)\Gamma \left(a_2\right)}+y^{-a_2}\frac{\Gamma \left(a_3\right)\Gamma \left(a_1-a_2\right)}{\Gamma \left(a_3-a_2\right)\Gamma \left(a_1\right)}\right].
\end{gather}

\subsection{Matching The Solutions} \label{sec:matching}
Additionally, one should consider $\{\Omega,\Omega_0\}\ll 1$ condition, which implies $\frac{1}{\Omega}\gg1$. In the region $\Omega y \ll1$, we can approximate \cite{xo,st},
\begin{align}
B_0 &\approx-l(l+1) ,\\
B_1 &\approx \left(\Omega-m_l \Omega_H\right)^2 \\
B_2 &\approx 0.
\end{align}
Additionally, in the region $ y\gg l(l+1)>1$, one obtains,
\begin{align}
A_0 &=(\Omega^2 - \Omega_0^2) \left[1 - \frac{4 (P^2 + Q^2)}{r_s^2}\right], \label{A0}\\
A_1 &\approx 0 ,\\
A_2 &\approx -l(l+1).
\end{align}

It is important to emphasize that the scalar field energy, encoded in the frequency $\Omega$, directly controls the size of the overlap (matching) region in which the AAM solution is valid. In particular, the extent of this region is not fixed geometrically, but depends sensitively on the energy scale of the perturbation.

To make this explicit, we examine the two limiting domains:
\begin{enumerate}
\item In the near-horizon region, the condition $\Omega y \ll 1$ translates into
\begin{equation}
r - r_+ \ll \frac{r_+}{\Omega}.
\label{111}
\end{equation}

\item In the far region, the requirement $y \gg 1$ becomes
\begin{equation}
r_+ \ll r - r_+.
\label{112}
\end{equation}
\end{enumerate}

Combining equations (\ref{111}) and (\ref{112}), we obtain the overlap (matched) region,
\begin{equation}
r_+ \ll r - r_+ \ll \frac{r_+}{\Omega}.
\label{AAMregion}
\end{equation}

This double inequality clearly shows that the width of the matching region is inversely proportional to $\Omega$. Therefore, smaller scalar frequencies enlarge the overlap domain, leading to a more reliable and accurate matching procedure.

It is also straightforward to show,
\begin{align}
-i\left(\sqrt{B_1}+\sqrt{B_2}\right)-a_1 &= -\beta-\frac{1}{2},\\
-i\left(\sqrt{B_1}+\sqrt{B_2}\right)-a_2 &= \beta-\frac{1}{2},
\end{align}
that yield the following relations,
\begin{gather}
N_{\infty 1}=N_{H1}\frac{\Gamma(a_3)\Gamma(a_1-a_2)}{\Gamma(a_3-a_2)\Gamma(a_1)}, \label{match1}\\
N_{\infty 2}=N_{H1}\frac{\Gamma(a_3)\Gamma(a_2-a_1)}{\Gamma(a_3-a_1)\Gamma(a_2)}. \label{match2}
\end{gather}

We now investigate the asymptotic behavior of the far-region solution \eqref{farsol}. In the limit $z\to\infty$, the confluent hypergeometric function $F_{1,1}$ can be transformed using the identity given in \eqref{faratfar}. 
We begin with the first term of \eqref{farsol},
\begin{multline}
N_{\infty 1}e^{\frac{1}{2}i\alpha z}z^{\frac{1}{2}+\beta}
F_{1,1}\left(\frac{1}{2}+\beta+i\frac{A_1}{\alpha},\,1+2\beta,\,i\alpha z\right)
= \\
N_{\infty 1}\Bigg[
(i\alpha)^{-\frac{1}{2}-\beta+i\frac{A_1}{\alpha}}
z^{i\frac{A_1}{\alpha}}e^{i\frac{\alpha}{2}z}
\frac{\Gamma(1+2\beta)}
{\Gamma\left(\frac{1}{2}+\beta+i\frac{A_1}{\alpha}\right)}
\\
+
(-i\alpha)^{-\frac{1}{2}-\beta-i\frac{A_1}{\alpha}}
z^{-i\frac{A_1}{\alpha}}e^{-i\frac{\alpha}{2}z}
\frac{\Gamma(1+2\beta)}
{\Gamma\left(\frac{1}{2}+\beta-i\frac{A_1}{\alpha}\right)}
\Bigg],
\end{multline}
while the second term expands as,
\begin{multline}
N_{\infty 2}e^{-\frac{1}{2}i\alpha z}z^{\frac{1}{2}-\beta}
F_{1,1}\left(\frac{1}{2}-\beta+i\frac{A_1}{\alpha},\,1-2\beta,\,i\alpha z\right)
= \\
N_{\infty 2}\Bigg[
(i\alpha)^{-\frac{1}{2}+\beta+i\frac{A_1}{\alpha}}
z^{i\frac{A_1}{\alpha}}e^{i\frac{\alpha}{2}z}
\frac{\Gamma(1-2\beta)}
{\Gamma\left(\frac{1}{2}-\beta+i\frac{A_1}{\alpha}\right)}
\\
+
(-i\alpha)^{-\frac{1}{2}+\beta-i\frac{A_1}{\alpha}}
z^{-i\frac{A_1}{\alpha}}e^{-i\frac{\alpha}{2}z}
\frac{\Gamma(1-2\beta)}
{\Gamma\left(\frac{1}{2}-\beta-i\frac{A_1}{\alpha}\right)}
\Bigg].
\end{multline}

Collecting the ingoing and outgoing contributions, the asymptotic far-field solution takes the compact form,
\begin{equation}
R_{far}(z\to\infty) = A_{in}\,e^{-\frac{1}{2}i\alpha z}\,z^{-i\frac{A_1}{\alpha}} + A_{out}\,e^{\frac{1}{2}i\alpha z}\,z^{i\frac{A_1}{\alpha}}, \label{faratfar}
\end{equation}
where
\begin{align}
A_{out} &= N_{\infty 1}(i\alpha)^{-\frac{1}{2}-\beta+i\frac{A_1}{\alpha}} \frac{\Gamma(1+2\beta)} {\Gamma\left(\frac{1}{2}+\beta+i\frac{A_1}{\alpha}\right)} + N_{\infty 2}(i\alpha)^{-\frac{1}{2}+\beta+i\frac{A_1}{\alpha}} \frac{\Gamma(1-2\beta)} {\Gamma\left(\frac{1}{2}-\beta+i\frac{A_1}{\alpha}\right)},
\label{Aout}
\\
A_{in} &= N_{\infty 1}(-i\alpha)^{-\frac{1}{2}-\beta-i\frac{A_1}{\alpha}} \frac{\Gamma(1+2\beta)} {\Gamma\left(\frac{1}{2}+\beta-i\frac{A_1}{\alpha}\right)}+ N_{\infty 2}(-i\alpha)^{-\frac{1}{2}+\beta-i\frac{A_1}{\alpha}} \frac{\Gamma(1-2\beta)} {\Gamma\left(\frac{1}{2}-\beta-i\frac{A_1}{\alpha}\right)}.
\label{Ain}
\end{align}

The constants $N_{\infty1,2}$ can be expressed in terms of the near-horizon coefficient $N_{H1}$ through \eqref{match1} and \eqref{match2}. 

\subsection{Superradiance Amplification}
Following the derivation in the previous section yielding \eqref{Zori}, the amplification factor is defined as, \cite{refcoef,Suphot,Brito},
\begin{equation}
Z_{l,m_l,a,P,Q} \equiv \left|\frac{A_{\mathrm{out}}}{A_{\mathrm{in}}}\right|^2 - 1 ,
\label{SuperAmp}
\end{equation}
and directly measures the fractional change in energy of the scattered mode relative to the incident flux.

Its interpretation follows from the energy flux at the horizon:
\begin{itemize}
\item $Z_{l,m_l,a,P,Q} > 0$ corresponds to superradiant amplification, implying that energy is extracted from the black hole,
\item $Z_{l,m_l,a,P,Q} < 0$ indicates net absorption,
\item $Z_{l,m_l,a,P,Q}=0$ at the critical frequency $\Omega = m_l \Omega_H$, where the horizon flux vanishes.
\end{itemize}

\begin{figure}[h]
    \centering
    \includegraphics[scale=1]{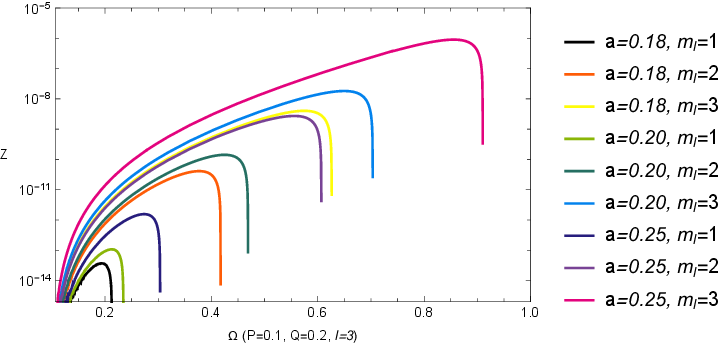}
    \caption{Amplification factor $Z_{l,m_l,a,P,Q}$ versus frequency $\Omega$ for several values of $l, m_l$ and $a$, with fixed $\Omega_0=0.1$ and $r_s=1$.}
    \label{Amplification}
\end{figure}

Figure \ref{Amplification} displays $Z_{l,m_l,a,P,Q}$ for a neutral bosonic field propagating in the dyonic Kerr-Sen background as a function of the frequency $\Omega$. We fix $\Omega_0 = 0.1$ and choose $r_s = 1$, $P = 0.1$ and $Q = 0.2$. Superradiance amplification occurs exclusively for co-rotating modes with $m_l > 0$, whose angular momentum is aligned with the black hole rotation. In this case, the condition,
\begin{equation}
\Omega < m_l \Omega_H,
\end{equation}
defines a finite frequency interval in which the horizon energy flux becomes negative, allowing rotational energy to be extracted. The amplification increases within this interval and vanishes at $\Omega = m_l \Omega_H$, which is the cut off of the corresponding mode. For co-rotating modes with $\Omega > m_l \Omega_H$, the flux reverses sign and absorption dominates. In the static limit $\Omega_H = 0$, the superradiant interval collapses entirely.

By contrast, counter-rotating modes ($m_l < 0$) never satisfy the superradiant condition. Since $\Omega - m_l \Omega_H$ remains positive for all $\Omega > 0$, the associated horizon flux is always inward and the amplification factor stays negative. Physically, these modes transfer angular momentum to the black hole rather than extracting it. 

For fixed $m_l$, increasing the horizon angular velocity $\Omega_H$ enlarges the superradiant window and enhances the maximal amplification, reflecting the greater rotational energy reservoir. Similarly, larger positive $m_l$ broaden the interval $\Omega < m_l \Omega_H$ and increase the peak value of $Z_{l,m_l,a,P,Q}$, indicating stronger coupling between higher co-rotating multipoles and the black hole rotation. While the presence of electric and magnetic charges slightly modifies the quantitative profile, the qualitative behavior remains consistent with that of the four-dimensional Kerr family \cite{Brito,Franzin,Senjaya:2025bbp}.

\begin{figure}[h]
    \centering
    \includegraphics[scale=1]{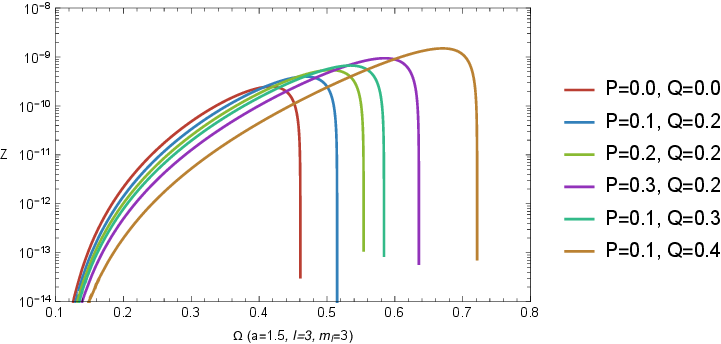}
    \caption{Amplification factor $Z_{l,m_l,a,P,Q}$ versus frequency $\Omega$ for several values of $P, Q$ with fixed $\Omega_0=0.1$ and $r_s=1, a=0.15, l=m_l=3$.}
    \label{Amplification1}
\end{figure}

In figure \ref{Amplification1}, we investigate the influence of black hole's charges on the amplification factor, fixing $a=0.15, r_s=1, \Omega_0=0.1$ and $l= m_l=3$ in all cases. Increasing either $P$ or $Q$ yields suppression to the superradiance with same effect due to the equation \eqref{A0}.

\begin{figure}[h]
    \centering
    \includegraphics[scale=1]{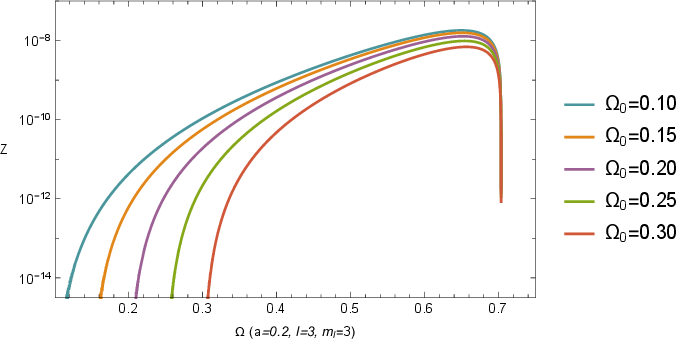}
    \caption{Amplification factor $Z_{l,m_l,a,P,Q}$ versus frequency $\Omega$ for several values of $\Omega_0$ with fixed $r_s=1, a=0.2, P=0.1, Q=0,2, l=m_l=3$.}
    \label{Amplification2}
\end{figure}

In figure \ref{Amplification2}, we examine the impact of the scalar field mass on the amplification factor. The black hole parameters are fixed to $r_s = 1$, $a = 0.2$, $P = 0.1$ and $Q = 0.2$, while the angular numbers are chosen as $l = m_l = 3$ throughout. The presence of a nonzero scalar mass introduces a kinematic threshold for propagation at spatial infinity. Only modes with,
\begin{equation}
\Omega > \Omega_0,
\end{equation}
can propagate as scattering states, while lower frequencies correspond to evanescent behavior. Superradiance therefore occurs within the restricted interval,
\begin{equation}
\Omega_0 < \Omega < m_l \Omega_H.
\end{equation}

Since the lower bound of this interval is set by the mass parameter $\Omega_0$, decreasing the scalar mass shifts the propagation threshold toward smaller frequencies and enlarges the available superradiant window. Consequently, lighter scalar fields exhibit broader and stronger amplification and typically achieve larger peak values of $Z_{l,m_l,a,P,Q}$.  In contrast, increasing the scalar mass reduces the width of the allowed frequency band and suppresses the maximal amplification. When $\Omega_0 \ge m_l \Omega_H$, the superradiant window closes entirely and amplification becomes impossible despite the presence of black hole rotation. It is important to note that the cutoff frequency doesn't change throughout the variation of the scalar mass.

\subsection{Greybody Factor}
Unlike an ideal black body enclosed in a cavity, a black hole does not emit a perfect Planck spectrum as measured at spatial infinity. Although Hawking radiation is thermally generated in the near-horizon region, the emitted quanta must propagate through the curved spacetime geometry before reaching a distant observer. During this propagation, they encounter an effective gravitational potential barrier that partially reflects the outgoing radiation. As a result, the spectrum observed at infinity is distorted relative to the original thermal emission. The frequency-dependent transmission probability associated with this scattering process is known as the greybody factor $\Gamma_{l,m_l,a,P,Q}$ \cite{Sharif:2022yxg,Javed:2024jty}. It encodes the imprint of spacetime geometry on the Hawking flux.

Physically, the greybody factor represents the probability that a mode of frequency $\Omega$ emitted near the horizon successfully reaches spatial infinity. By time-reversal symmetry, it is equivalently the absorption probability for an incoming wave of the same frequency to penetrate the potential barrier and reach the horizon \cite{Harmark:2007jy,Hyun:2019qgq}. In terms of the asymptotic wave amplitudes, it can be written as,
\begin{equation}
\Gamma_{l,m_l,a,P,Q} \equiv 1 - \left| \frac{A_{\mathrm{out}}}{A_{\mathrm{in}}} \right|^2 .
\end{equation}

Once the near-horizon and asymptotic solutions of the radial equation are obtained, the greybody factor follows directly from equations \eqref{Ain} and \eqref{Aout}.

We now examine the low-frequency behaviour of the greybody factor $\Gamma_{l,m_l,a,P,Q}$ in the rotating dyonic Kerr-Sen spacetime. The combined effect of rotation and electric/magnetic charges modifies the effective radial potential experienced by scalar perturbations, thereby altering their transmission probability from the near-horizon region to spatial infinity.

\begin{figure}[h]
    \centering
        \includegraphics[scale=0.6]{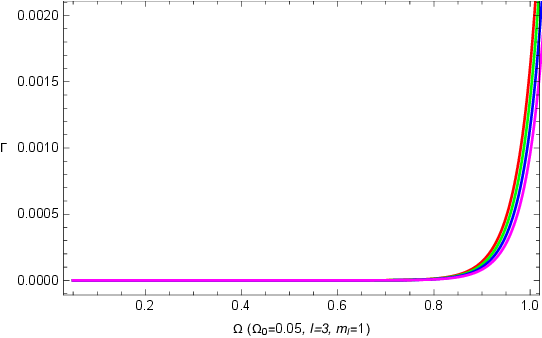}
        \includegraphics[scale=0.63]{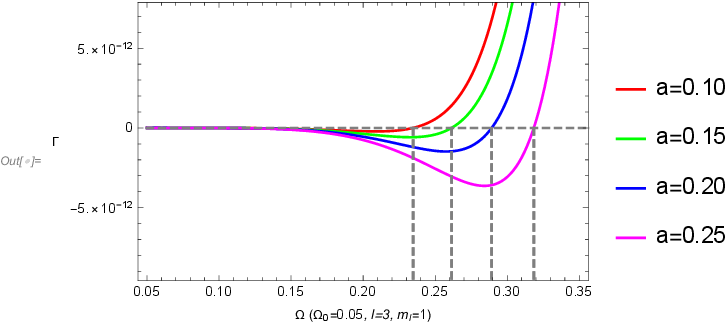}
    \caption{Greybody factor profile of the $l=3, m_l=1$ mode for various values of the rotation parameter $a$, with $\Omega_0=0.05, r_s=1, P=0.1$ and $Q=0.2$ fixed.}
    \label{coRGBS1}
\end{figure}

In figure \cref{coRGBS1}, we display the greybody factor for the mode $l=3$, $m_l=1$ as a function of the frequency $\Omega$, for several values of the rotation parameter $a$. Throughout the analysis we fix the scalar mass $\Omega_0 = 0.05$ and the black hole parameters $r_s = 1$, $P = 0.1$ and $Q = 0.2$. For co-rotating modes ($m_l \geq 1$), a superradiant regime emerges within the frequency interval,
\begin{equation}
\Omega_0 < \Omega < m_l \Omega_H ,
\end{equation}
where the greybody factor becomes negative. Since,
\begin{equation}
\Gamma_{l,m_l,a,P,Q} = 1 - \left| \frac{A_{\mathrm{out}}}{A_{\mathrm{in}}} \right|^2 ,
\end{equation}
a negative value implies $\left|A_{\mathrm{out}}\right| > \left|A_{\mathrm{in}}\right|$, corresponding to superradiant amplification. In this regime, the horizon flux is negative and rotational energy is extracted from the black hole.

As the rotation parameter $a$ increases, the magnitude of the negative minimum grows and the superradiant interval broadens. This behaviour reflects the larger horizon angular velocity $\Omega_H$ and the enhanced rotational energy reservoir available for extraction. Faster rotation therefore leads to stronger amplification and a more pronounced deviation from purely absorptive behaviour. By contrast, for counter-rotating modes ($m_l \leq 0$), the condition $\Omega < m_l \Omega_H$ cannot be satisfied for positive frequencies. Consequently, the greybody factor remains positive throughout the spectrum, indicating that the flux across the horizon is always inward and no superradiant amplification occurs. In this sector, scalar waves are partially absorbed for all frequencies.

\begin{figure}[h]
    \centering
        \includegraphics[scale=1]{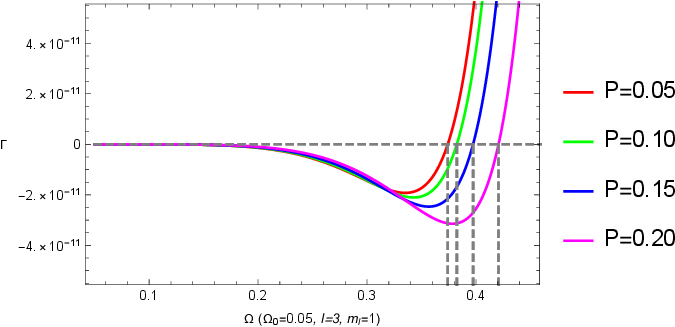}
    \caption{Greybody factor profile of the $l=3, m_l=1$ mode for various values of the dyonic charge $P$, with $\Omega_0=0.05, r_s=1, Q=0.2$ and $a=0.3$ fixed.}
    \label{coRGBS2}
\end{figure}

In Fig.~\cref{coRGBS2}, we present the greybody factor for the mode $l=3$, $m_l=1$ as a function of the frequency $\Omega$, for several values of the magnetic charge $P$. Varying the electric charge produces an analogous effect. Throughout the analysis we fix the scalar mass $\Omega_0 = 0.05$ and the black hole parameters $r_s = 1$, $a = 0.3$ and $Q = 0.2$. As the dyonic charge increases, the superradiant amplification becomes more pronounced. In particular, the magnitude of the negative minimum of $\Gamma_{l,m_l,a,P,Q}$ increases and the superradiant interval $\Omega_0 < \Omega < m_l \Omega_H$ broadens. We note that the presence of electric and magnetic charges modifies both the horizon structure and thereby affecting the horizon angular velocity $\Omega_H$ and the efficiency of rotational energy extraction, in the same way due to the equation \eqref{A0}. Consequently, larger dyonic charges enhance the deviation from purely absorptive behaviour and lead to stronger amplification in the co-rotating sector.

By contrast, for counter-rotating modes ($m_l \leq 0$), the condition $\Omega < m_l \Omega_H$ cannot be satisfied for positive frequencies. The greybody factor therefore remains positive across the entire spectrum, indicating purely absorptive scattering with no superradiant amplification.

\section{Superradiance Energy Extraction}
In this section we investigate the energy extracted from the black hole through superradiant scattering. We begin with the outgoing energy flux measured by an observer at spatial infinity as derived in Appendix \ref{E},
\begin{align}
\dot{E}_{\rm out} = \frac{\omega k_\infty}{2} |\mathcal{R}|^2.
\end{align}

For simplicity, and without loss of generality for the discussion of superradiant amplification, we restrict to the massless case, $\Omega_0=0$, so that the asymptotic dispersion relation reduces to $k_\infty= \Omega/r_s= \omega$.

In a realistic scattering process, the incident scalar field is not strictly monochromatic but is described by a wave packet composed of modes with different frequencies. The total extracted energy is therefore obtained by integrating over the spectral distribution of the incident field,
\begin{align} 
\dot{E}_{\rm tot} = \int_0^\infty \frac{\omega^2}{2} |\mathcal{R}(\omega)-1|^2 \, n(\omega)\, d\omega =\int_0^\infty \frac{\omega^2}{2}Z_{l,m_l,a,P,Q} \, n(\omega)\, d\omega,
\end{align}
where $n(\omega)$ represents the normalized spectral weight of the incoming radiation.

If the incident field is in thermal equilibrium at temperature $T_{\rm tem}$, the occupation number per mode follows the Bose–Einstein distribution,
\begin{align}
\tilde n(\omega) = \frac{1}{e^{\omega/T_{\rm tem}} - 1}. \end{align}

To use the Bose–Einstein occupation number as a normalized spectral weight, one must take into account the density of states in three spatial dimensions. For a massless scalar field, the number of modes in the momentum interval $(k, k+dk)$ is proportional to $k^2 dk$. Since $k=\omega$ for massless modes, the phase-space measure becomes $\omega^2 d\omega$.

Therefore, the properly normalized spectral weight is defined as,
\begin{align}
n(\omega) = \frac{\tilde n(\omega)\,\omega^2} {\displaystyle \int_0^\infty \tilde n(\omega)\,\omega^2 d\omega},
\end{align}
where,
\begin{align}
\tilde n(\omega) = \frac{1}{e^{\omega/T_{\rm tem}} - 1}
\end{align}
is the Bose–Einstein occupation number per mode.

The denominator evaluates to,
\begin{align} 
\int_0^\infty \frac{\omega^2}{e^{\omega/T_{\rm tem}} - 1} \, d\omega = 2\,\zeta(3)\,T_{\rm tem}^3,
\end{align}
so that,
\begin{align}
n(\omega) = \frac{1}{2\zeta(3)T_{\rm tem}^3}\left( \frac{\omega^2}{e^{\omega/T_{\rm tem}} - 1}\right).
\end{align}

Here $\zeta(3)\approx 1.202$ is the Riemann zeta function. 
This normalization ensures,
\begin{align}
\int_0^\infty n(\omega)\, d\omega = 1.
\end{align}

It is important to emphasize that the reflection amplitude $|\mathcal{R}(\omega)|^2$ is entirely determined by the black hole geometry and does not depend on $T_{\rm tem}$. 
The temperature only determines the population of the incident modes. To isolate the superradiant energy extraction from the intrinsic Hawking emission, we consider $T_{\rm tem} \gg T_H$, so that the incoming flux dominates over the spontaneous Hawking flux. In this regime, the net energy gain arises purely from stimulated superradiant scattering \cite{Wondrak_2018}.

\begin{figure}[h]
    \centering
        \includegraphics[scale=0.9]{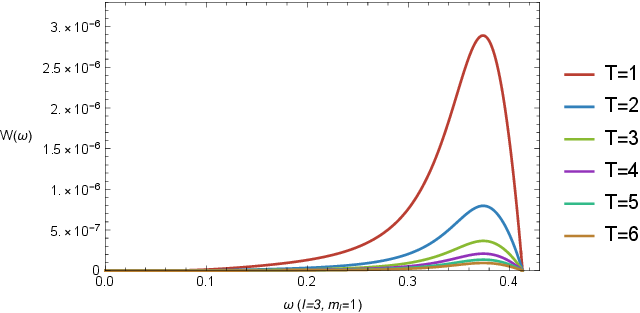}\\
         \includegraphics[scale=0.9]{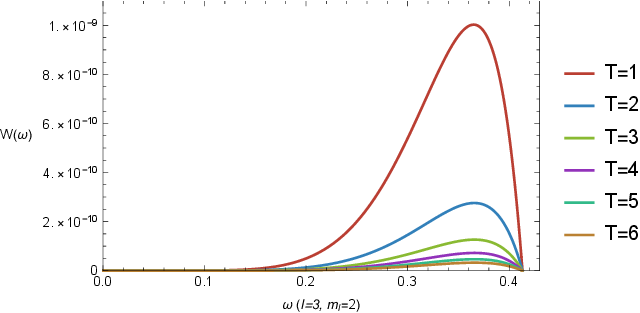}
    \caption{Energy extraction rate profile of the $l=3, m_l=1,2$ modes for various values of background temperature, with $r_s=1, P=0.3, Q=0.2$ and $a=0.1$ fixed.}
    \label{Spec}
\end{figure}

In figure \ref{Spec} we plot the spectral energy extraction rate,
\begin{equation}
W(\omega)=\frac{d\dot{E}_{\rm out}}{d\omega}4\zeta(3) =\frac{Z_{l,m_l,a,P,Q} }{T_{\rm tem}^3} \left( \frac{\omega^4}{e^{\omega/T_{\rm tem}} - 1} \right),
\end{equation}
as a function of the frequency $\omega$. 

The spectrum displays a thermally weighted profile shaped by the superradiant amplification factor $Z_{l,m_l,a,P,Q}$. For all temperatures, energy extraction remains strictly confined to the superradiant band and vanishes at the critical frequency $\omega = m_l\Omega_H$. As the background temperature $T=T_{\rm tem}$ increases, the overall amplitude of the spectrum rises markedly, while the peak becomes higher and broader. Physically, increasing the temperature enhances the incident bosonic flux, thereby amplifying the stimulated superradiant contribution. We also observe that lower multipole numbers $m_l$ lead to significantly stronger energy extraction, indicating a more efficient coupling between low-$m_l$ modes and the rotational energy of the black hole.

\section{Conclusions}
In this work, we present a comprehensive analytic study of superradiant scattering of a massive scalar field in the dyonic Kerr-Sen black-hole background. Starting from the Klein-Gordon equation, we exploit the separability of the spacetime and recast the radial equation into a Schrödinger-like form via the tortoise coordinate. This formulation makes the conserved flux structure explicit and allows a transparent identification of reflection and transmission coefficients. A direct near-horizon analysis reveals that the sign of the horizon momentum governs the energy flow:, i.e. from the relation $|\mathcal{R}|^{2} = 1 - \frac{\Omega - \Omega_{H}}{\sqrt{\Omega^{2}-\Omega_{0}^{2}}}\,|\mathcal{T}|^{2}$, amplification occurs whenever $\Omega < \Omega_H$, where the horizon flux becomes negative and rotational energy is extracted, while higher frequencies correspond to ordinary absorption.

To go beyond the threshold condition and obtain the full frequency dependence, we implement the analytical asymptotic matching method. The radial equation simplifies dramatically in the near-horizon and asymptotic regions, where the solutions are expressed in terms of ${}_2F_1$ and ${}_1F_1$ hypergeometric functions, respectively. By analytically continuing both solutions into the common overlap region $r_+ \ll r-r_+ \ll r_+/\Omega$ and matching their asymptotic expansions, we determine the relative normalization of ingoing and outgoing modes. 

We find that the amplification factor $Z_{l,m_l,a,P,Q}$ is strictly positive only for co-rotating modes ($m_l>0$), and only within the finite superradiant band $\Omega_0 < \Omega < m_l\Omega_H$. At the critical frequency $\Omega = m_l\Omega_H$, the amplification vanishes exactly, marking the transition between energy extraction and pure absorption. Increasing the black hole spin or its electric and magnetic charges enlarges this superradiant window and raises the peak amplification, demonstrating that rotational energy extraction becomes more efficient in more rapidly rotating and more highly charged backgrounds. Higher multipole numbers further strengthen the coupling between the scalar perturbation and the black hole rotation, leading to enhanced amplification for selected modes.

The scalar mass introduces a fundamental propagation threshold $\Omega>\Omega_0$, which constrains the accessible frequency range and can substantially weaken the superradiant effect. As $\Omega_0$ approaches the upper bound $m_l\Omega_H$, the allowed band progressively narrows and ultimately collapses, completely suppressing amplification. In precise correspondence, the greybody factor becomes negative in exactly the same interval, confirming that superradiance occurs when the amplification factor is positive. By contrast, counter-rotating modes ($m_l<0$) remain purely absorptive throughout the entire spectrum, reflecting the directional nature of rotational energy extraction.

We also investigate the total energy extracted from the dyonic Kerr--Sen black hole by considering a massless scalar field with a thermal incident spectrum. In the high-temperature regime $T_{\rm tem} \gg T_H$, the incoming flux overwhelmingly dominates over spontaneous Hawking emission, so that the extracted energy is governed primarily by stimulated superradiance. Our results show that the total extracted power increases rapidly with $T_{\rm tem}$, i.e. hotter environment, with both the overall magnitude and the spectral peak growing as the temperature rises. In addition, lower multipole numbers $m_l$ yield significantly larger energy extraction, indicating a more efficient coupling of low-$m_l$ co-rotating modes to the rotational energy of the black hole. 

Finally, our results suggest potential phenomenological implications. Since both the superradiant threshold and the amplification factor depend explicitly on the electric and magnetic charges, the superradiant spectrum carries information about the underlying dilaton–axion structure of the geometry. Deviations in energy extraction rates or greybody factors from their Kerr counterparts could therefore serve as indirect signatures of string-inspired charge sectors, particularly in scenarios where wave scattering processes are observationally relevant. Future work may extend this analysis beyond the small-$(a,P,Q)$ regime to explore the fully nonlinear parameter space, where new effects could arise. It would also be interesting to consider higher-spin perturbations, such as electromagnetic and gravitational fields, in order to test the universality of these charge-induced modifications in black-hole wave dynamics.

\bibliography{sn-bibliography}

\begin{appendices} 
    \include{appendix} 
\section{Normal Form}\label{AppendixA}
The normal form of a second–order ordinary differential equation is obtained by removing the first–derivative term so that the equation is written solely in terms of the highest derivative \cite{NIST}. 

Consider the linear equation,
\begin{equation}
\frac{d^2 y}{dx^2}+p(x)\frac{dy}{dx}+q(x)y=0.
\label{generalODE}
\end{equation}

To eliminate the first–order term, we introduce the transformation,
\begin{equation}
y(x)=Y(x)\exp\!\left(-\frac{1}{2}\int p(x)\,dx\right),
\end{equation}
which is constructed precisely so that the resulting equation for $Y(x)$ contains no $dY/dx$ term \cite{2420}. After straightforward substitution, equation \eqref{generalODE} reduces to,
\begin{equation}
\frac{d^2 Y}{dx^2} +\left( -\frac{1}{2}\frac{dp}{dx} -\frac{1}{4}p^2 +q \right)Y=0.
\label{normalform}
\end{equation}

Defining,
\begin{equation}
Q(x)=-\frac{1}{2}\frac{dp}{dx}-\frac{1}{4}p^2+q,
\end{equation}
the equation assumes the compact Schrödinger-like form,
\begin{equation}
\frac{d^2 Y}{dx^2}=-Q(x)Y.
\end{equation}

This representation is particularly useful for qualitative analysis. If $Q(x)>0$, the equation is locally oscillatory and the solution necessarily crosses the $x$-axis. Moreover, if
\begin{equation}
    \int_{1}^{\infty}Q(x)\,dx=\infty,
\end{equation}
then $Y(x)$ possesses infinitely many zeros on the positive axis \cite{2420}. In contrast, if $Q(x)<0$, the solution is non-oscillatory and admits at most one zero.

\section{The Confluent Heun Equation and Its Solutions}\label{AppendixB}
The confluent Heun equation arises whenever a second–order differential equation possesses two regular singular points and one irregular singular point at infinity. Its canonical form reads \cite{Heun},
\begin{equation}
\frac{d^2\psi_H}{dx^2}+\left(\alpha+\frac{\beta+1}{x}+\frac{\gamma+1}{x-1}\right)\frac{d\psi_H}{dx}+\left(\frac{\mu}{x}+\frac{\nu}{x-1}\right)\psi_H=0,
\label{HeunCanonical}
\end{equation}
with parameters,
\begin{align}
\mu &= \frac{1}{2}\left(\alpha-\beta-\gamma+\alpha\beta-\beta\gamma\right)-\eta, \\ \nu &= \frac{1}{2}\left(\alpha+\beta+\gamma+\alpha\gamma+\beta\gamma\right) +\delta+\eta .
\end{align}

Equation \eqref{HeunCanonical} admits two linearly independent local solutions expressed through the confluent Heun function,
\begin{equation}
\psi_H = A\,\operatorname{HeunC}(\alpha,\beta,\gamma,\delta,\eta,x) + B\,x^{-\beta} \operatorname{HeunC}(\alpha,-\beta,\gamma,\delta,\eta,x),
\label{HeunGeneral}
\end{equation}
where $A$ and $B$ are arbitrary constants. 

A remarkable feature of the confluent Heun function is that it reduces to a polynomial of degree $n_r$ provided the parameter constraint
\begin{equation}
\frac{\delta}{\alpha} +\frac{\beta+\gamma}{2} +1 = - n_r, \qquad n_r\in\mathbb{Z},
\label{HeunPolynomial}
\end{equation}
is satisfied. This condition plays a central role in spectral problems where regularity and normalizability select discrete solutions.

To expose the underlying structure of equation \eqref{HeunCanonical}, we transform it into its normal form following Appendix \ref{AppendixA}. Identifying,
\begin{align}
p(x) &= \alpha+\frac{\beta+1}{x}+\frac{\gamma+1}{x-1}, \\
q(x) &= \frac{\mu}{x}+\frac{\nu}{x-1},
\end{align}
we introduce,
\begin{equation}
\psi_H(x) = \Psi_H(x) e^{-\frac{1}{2}\alpha x} x^{-\frac{1}{2}(\beta+1)} (x-1)^{-\frac{1}{2}(\gamma+1)}.
\label{HeunTransform}
\end{equation}

Upon substitution, the first–derivative term is eliminated and the equation assumes the Schrödinger-like form,
\begin{equation}
\frac{d^2\Psi_H}{dx^2}+V_{\text{eff}}(x)\,\Psi_H=0,
\label{HeunNormal}
\end{equation}
where the effective potential is,
\begin{equation}
V_{\text{eff}}(x)=-\frac{\alpha^2}{4}+\frac{\frac{1}{2}-\eta}{x}+\frac{\frac{1}{4}-\frac{\beta^2}{4}}{x^2}+\frac{-\frac{1}{2}+\delta+\eta}{x-1}+\frac{\frac{1}{4}-\frac{\gamma^2}{4}}{(x-1)^2}.
\end{equation}

The structure of $V_{\text{eff}}(x)$ makes the singular behaviour at $x=0$ and $x=1$ manifest, while the constant term $-\alpha^2/4$ controls the asymptotic growth at infinity. Reconstructing $\Psi_H$ in terms of Heun functions yields,
\begin{multline}
\Psi_H = e^{\frac{1}{2}\alpha x} x^{\frac{1}{2}(\beta+1)} (x-1)^{\frac{1}{2}(\gamma+1)} \times \\ \left[ A\,\operatorname{HeunC}(\alpha,\beta,\gamma,\delta,\eta,x) + B\,x^{-\beta} \operatorname{HeunC}(\alpha,-\beta,\gamma,\delta,\eta,x) \right].
\end{multline}

\section{The Confluent Hypergeometric Equation}\label{AppendixC}
The confluent hypergeometric equation appears as a canonical normal form for a wide class of radial and wave-type problems. It can be written as \cite{Bell},
\begin{equation}
\frac{d^2\psi_C}{dx^2}+\left(-\frac{1}{4}+\frac{k}{x}+\frac{\frac{1}{4}-m^2}{x^2}\right)\psi_C=0.
\label{WhittakerNormal}
\end{equation}

Equation \eqref{WhittakerNormal} is recognized as the Whittaker form of the confluent hypergeometric equation. Its two linearly independent solutions are expressed in terms of Whittaker functions,
\begin{equation}
\psi_C=A\,M_{k,m}(x)+B\,M_{k,-m}(x),
\end{equation}
where $A$ and $B$ are arbitrary constants.

The Whittaker functions are directly related to the confluent hypergeometric function ${}_1F_1(a,b,x)$ through
\begin{equation}
M_{k,\pm m}(x) = x^{\frac{1}{2}\pm m} e^{-\frac{x}{2}} {}_1F_1 \left( \frac{1}{2}-k\pm m, 1\pm 2m, x \right).
\end{equation}

Note that the power-law prefactor governs the behaviour near the origin, while the exponential term controls the large-$x$ growth. In the asymptotic regime $x\to\infty$, the confluent hypergeometric function admits the expansion \cite{NIST},
\begin{multline}
{}_1F_1(a,b,|x|\to\infty)=\frac{\Gamma(b)}{\Gamma(a)}e^{x}x^{a-b}{}_2F_0\left(b-a,1-a;-\frac{1}{x}\right)\\+\frac{\Gamma(b)}{\Gamma(b-a)}(-x)^{-a}{}_2F_0\left(a,a-b+1;-\frac{1}{x}\right),
\label{HyperAsymptotic}
\end{multline}
where ${}_2F_0(a,b;x)$ denotes the generalized hypergeometric function of type $\{2,0\}$.

\section{The Gauss Hypergeometric Equation}\label{AppendixD}
The Gauss hypergeometric equation is the prototypical second–order differential equation with three regular singular points. Its canonical form is \cite{Bell},
\begin{equation}
x(1-x)\frac{d^2\psi_G}{dx^2}+\left[a_3-(a_1+a_2+1)x\right]\frac{d\psi_G}{dx}-a_1a_2\psi_G=0.
\label{GaussCanonical}
\end{equation}

Equation \eqref{GaussCanonical} admits two linearly independent solutions expressed in terms of the Gauss hypergeometric function ${}_2F_1$,
\begin{equation}
\psi_G=A\,{}_2F_1(a_1,a_2,a_3,x)+B\,x^{1-a_3}{}_2F_1(a_1-a_3+1,a_2-a_3+1,2-a_3,x),
\end{equation}
where $A$ and $B$ are arbitrary constants.

Following Appendix \ref{AppendixA}, the equation can be cast into normal form. Introducing,
\begin{equation}
\Psi_G=x^{\frac{a_3}{2}}(1-x)^{\frac{1}{2}(1+a_1+a_2-a_3)}\psi_G,
\end{equation}
equation \eqref{GaussCanonical} becomes,
\begin{equation}
\frac{d^2\Psi_G}{dx^2}-\left[\frac{x^2\!\left((a_1-a_2)^2-1\right)-2x\!\left((a_1+a_2-1)a_3-2a_1a_2\right)+a_3(a_3-2)}{4x^2(1-x)^2}\right]\Psi_G=0.
\label{GaussNormal}
\end{equation}

In this representation, the singular structure at $x=0$ and $x=1$ becomes explicit, while the numerator encodes the interplay between the parameters $a_1$, $a_2$ and $a_3$.

Now, consider now the differential equation,
\begin{equation}
x(1+x)\frac{d^2\psi_G}{dx^2}+(2x+1)\frac{d\psi_G}{dx}+\left(\frac{\mathcal{A}}{x}-\frac{\mathcal{B}}{x+1}+\mathcal{C}\right)\psi_G=0.
\label{GaussModified}
\end{equation}

Equation \eqref{GaussModified} can be mapped to the Gauss hypergeometric equation by a suitable change of variables, leading to the general solution,
\begin{multline}
\psi_G=(x+1)^{-i\sqrt{\mathcal{B}}}\Big[A\,x^{i\sqrt{\mathcal{A}}}{}_2F_1(\mathcal{A}_1,\mathcal{A}_2,\mathcal{A}_3,-x)\\+B\,x^{-i\sqrt{\mathcal{A}}}{}_2F_1(\mathcal{A}_4,\mathcal{A}_5,\mathcal{A}_6,-x)\Big],
\end{multline}
where,
\begin{align}
\mathcal{A}_1 &= \frac{1}{2}\left(1+\sqrt{1-4\mathcal{C}}\right)
+i\left(\sqrt{\mathcal{A}}-\sqrt{\mathcal{B}}\right), \\
\mathcal{A}_2 &= \frac{1}{2}\left(1-\sqrt{1-4\mathcal{C}}\right)
+i\left(\sqrt{\mathcal{A}}-\sqrt{\mathcal{B}}\right), \\
\mathcal{A}_3 &= 1+2i\sqrt{\mathcal{A}}, \\
\mathcal{A}_4 &= \frac{1}{2}\left(1+\sqrt{1-4\mathcal{C}}\right)
-i\left(\sqrt{\mathcal{A}}+\sqrt{\mathcal{B}}\right), \\
\mathcal{A}_5 &= \frac{1}{2}\left(1-\sqrt{1-4\mathcal{C}}\right)
-i\left(\sqrt{\mathcal{A}}+\sqrt{\mathcal{B}}\right), \\
\mathcal{A}_6 &= 1-2i\sqrt{\mathcal{A}}.
\end{align}

A key property of the Gauss hypergeometric function is its analytic continuation to large argument. The connection formula reads,
\begin{multline}
{}_2F_1(a_1,a_2,a_3,x)=\frac{\Gamma(a_2-a_1)\Gamma(a_3)}{\Gamma(a_2)\Gamma(a_3-a_1)}(-x)^{-a_1}{}_2F_1\!\left(a_1,a_1-a_3+1,a_1-a_2+1,\frac{1}{x}\right)\\+\frac{\Gamma(a_1-a_2)\Gamma(a_3)}{\Gamma(a_1)\Gamma(a_3-a_2)}(-x)^{-a_2}{}_2F_1\!\left(a_2,a_2-a_3+1,a_2-a_1+1,\frac{1}{x}\right).
\label{GaussConnection}
\end{multline}

\section{Derivation of the energy flux expression}\label{E}
In this appendix we derive the expression for the energy flux of the scalar field at spatial infinity. We begin with the Lagrangian density for a complex scalar field in curved spacetime,
\begin{align}
\mathcal{L} = g^{\mu\nu}\partial_{(\mu}\psi^*\partial_{\nu)}\psi -\frac{1}{2} m^2 |\psi|^2 ,
\end{align},
where
\begin{align}
\partial_{(\mu}\psi^*\partial_{\nu)}\psi=\frac{1}{2}\left(\partial_\mu\psi^*\partial_\nu\psi+\partial_\nu\psi^*\partial_\mu\psi\right).
\end{align}

The energy flux per unit solid angle measured at spatial infinity is defined as,
\begin{align}
\frac{d^2E}{dt\, d\Omega} = \lim_{r \to \infty} r^2 T^r_{\ t},
\end{align}
where $T_{\mu\nu}$ is the energy-momentum tensor of the scalar field,
\begin{align}
T^\mu_{\ \nu} = g^{\mu\lambda}\partial_{(\lambda}\psi^*\partial_{\nu)}\psi - \delta^\mu_{\ \nu}\mathcal{L}.
\label{T}
\end{align}

From the main text, the scalar field admits the separated solution,
\begin{align}
\psi = e^{-i\omega t + i m_l \phi} S^{m_l}_l(c_a,\cos\theta) R(r).
\end{align}

At spatial infinity, the radial function behaves as \eqref{Rastinf},
\begin{align}
R(r\to\infty) \approx \frac{1}{r} \left( R_{\rm inc} e^{-i k_\infty r} + R_{\rm ref} e^{i k_\infty r} \right),
\end{align}
where $k_\infty = \frac{\sqrt{\Omega^2 - \Omega_0^2}}{r_s}$.

We now compute the mixed component $T^r_{\ t}$. 
Using equation \eqref{T} and noting that asymptotically $g^{rr} (r\to\infty)\approx 1$, we obtain,
\begin{align}
T^r_{\ t} = \frac{1}{2} g^{rr} \left( \partial_r \psi^* \partial_t \psi + \partial_t \psi^* \partial_r \psi \right).
\end{align}

Using $\partial_t \psi = - i \omega \psi$ and $\partial_r \psi = e^{-i\omega t + i m_l \phi}S^{m_l}_l R'(r)$, this becomes,
\begin{align}
T^r_{\ t} = \frac{i\omega}{2} g^{rr} \left( R R'^* - R^* R' \right) |S^{m_l}_l|^2 = -g^{rr} \operatorname{Im}(RR'^*) |S^{m_l}_l|^2.
\end{align}

Substituting the asymptotic form of $R(r)$, the outgoing contribution at leading order in $1/r$ is obtained as follows,
\begin{align}
T^r_{\ t}(r\to\infty) \approx \frac{\omega k_\infty}{r^2} |R_{\rm ref}|^2 |S^{m_l}_l|^2 .
\end{align}

Finally, multiplying by $r^2$ and integrating over the solid angle using the normalization of the spheroidal harmonics,
\begin{align}
\int |S^{m_l}_l|^2 d\Omega = 1,
\end{align}
we obtain the outgoing energy flux,
\begin{align}
\dot{E}_{\rm out} = \frac{\omega k_\infty}{2} |R_{\rm ref}|^2=\frac{\omega k_\infty}{2} |\mathcal R|^2,
\end{align}
where we have chosen $|R_{\rm inc}|^2=1$ and $|\mathcal R|^2$ is the reflection probability defined as,
\begin{equation}
|\mathcal R|^2=\frac{|R_{\rm ref}|^2}{|R_{\rm inc}|^2}.
\end{equation}

\end{appendices}

\end{document}